\documentstyle[aps,psfig,multicol]{revtex}

\begin{document}

\draft

\title{Minimal relativity and $^3S_1$-$^3D_1$ pairing in symmetric 
nuclear matter}
 
\author{\O.\ Elgar\o y$^a$, L.\ Engvik$^a$, M.\ Hjorth-Jensen$^b$ and 
E.\ Osnes{$^a$}}

\address{$^a$Department of Physics, University of Oslo, N-0316 Oslo, Norway}

\address{$^b$Nordita, Blegdamsvej 17, DK-2100 K\o benhavn \O, Denmark}

\maketitle

\begin{abstract}
We present solutions of the coupled, non-relativistic
 $^3S_1$-$^3D_1$ gap equations for neutron-proton pairing 
in symmetric nuclear matter, 
and estimate relativistic effects by solving the same 
gap equations modified according to minimal relativity and using 
single-particle energies from a Dirac-Brueckner-Hartree-Fock calculation. 
As a main result we find that relativistic effects decrease the 
value of the gap 
at the saturation density $k_F=1.36\;{\rm fm}^{-1}$ considerably, 
in conformity with the lack of evidence for strong neutron-proton 
pairing in finite nuclei.  
\end{abstract}


\begin{multicols}{2}

The size of the neutron-proton (np) $^3S_1$-$^3D_1$ energy gap in 
symmetric nuclear matter has been a much debated issue since the 
first calculations of this quantity appeared.  While 
solutions of the BCS equations with bare nucleon-nucleon (NN) forces 
give a large energy gap of several MeV's at the saturation density 
$k_F=1.36\;{\rm fm}^{-1}$ ($\rho=0.17\;{\rm fm}^{-3}$) 
\cite{alm90,vonder91,tak93,baldo95},  
there is little empirical evidence from 
finite nuclei for such strong np pairing correlations.  
One possible resolution of this problem lies in the fact 
that all these calculations have neglected contributions from the so-called 
induced interaction.  Fluctuations in the isospin and the spin-isospin 
channel will probably make the pairing interaction more repulsive, 
leading to a substantially lower energy gap \cite{pethick97}.  
One often neglected aspect is that all non-relativistic calculations 
of the nuclear matter equation of state (EOS) with two-body NN forces 
fitted to scattering data fail to reproduce the empirical saturation 
point, seemingly regardless of the sophistication of the 
many-body scheme employed. 
For example, a Brueckner-Hartree-Fock (BHF) calculation of the 
EOS with one of the 
Bonn potentials would typically give saturation at $k_F=1.6$-$1.8\;{\rm fm}
^{-1}$.  In a non-relativistic approach it seems necessary to invoke 
three-body forces to obtain saturation at the empirical equilibrium 
density.  This leads one to be cautious when talking about pairing at 
the empirical nuclear matter saturation density when the energy gap 
is calculated within a pure two-body force model, as this density 
will be below the calculated saturation density for this two-body force, 
and thus one is calculating the gap at a density where the 
system is theoretically unstable.   
One even runs the risk, as pointed out in 
Ref.\ \cite{jack83}, that 
the compressibility is negative at the empirical saturation density, 
which means that the system is unstable against collapse into a 
non-homogeneous phase.  
A three-body force need not have dramatic consequences 
for pairing, which after all is a two-body phenomenon, but still it 
would be of interest to know what the $^3S_1$-$^3D_1$ gap is in 
a model in which the saturation properties of nuclear matter are reproduced.  
If one abandons a non-relativistic description, the empirical saturation 
point can be obtained within the Dirac-Brueckner-Hartree-Fock (DBHF) approach, 
as first pointed out by Brockmann and Machleidt \cite{brock90}.    
This might be fortuitous, since, among other things,  
important many-body effects are neglected 
in the DBHF approach.  Nevertheless, we found it interesting to investigate 
$^3S_1$-$^3D_1$ pairing in this model and compare our results with a  
corresponding non-relativistic calculation.

The first ingredient in our calculation is the self-consistent
evaluation of single-particle energies in symmetric nuclear matter 
starting from the meson-exchange potential models
of Machleidt and co-workers \cite{mac89}.  
For the non-relativistic (NR) calculations we use the 
BHF method, while the DBHF scheme is used in 
the relativistic (R) calculation.  Details of both approaches 
are found in Refs.\ \cite{brock90,cs86,hko95}.  Since BHF and DBHF 
are computationally very similar, we will here content ourselves 
with giving a brief description of the DBHF method. 
In this scheme, the single-particle energies and binding energy of 
nuclear matter is obtained   
using a medium renormalized NN potential $G$ defined through
the solution of the $G$-matrix equation 
\begin{equation}
       G(\omega)=V+VQ\frac{1}{\omega - QH_0Q}QG(\omega ),
       \label{eq:bg}
\end{equation}
where $\omega$ is the unperturbed energy of the interacting  nucleons,
$V$ is the free NN potential, $H_0$ is the unperturbed energy of the
intermediate scattering states,
and $Q$ is the Pauli
operator preventing scattering into occupied states.
Only ladder diagrams with two-particle intermediate states are included
in Eq.\ (\ref{eq:bg}).
In this work we solve Eq.\ (\ref{eq:bg}) using the Bonn A potential
defined in Table A.2 of Ref.\ \cite{mac89}. This potential model 
employs the Thompson \cite{brock90,thompson70} reduction of the
Bethe-Salpeter equation, and is tailored for relativistic
nuclear structure calculations.  For the non-relativistic calculation 
we employ the Bonn A potential with parameters from Table A.1 in 
Ref.\ \cite{mac89}.  This model employs the 
Blankenbecler-Sugar (BbS) reduction of the Bethe-Salpeter equation, and is 
therefore suited for non-relativistic calculations.  
For further details, see Refs.\ \cite{brock90,mac89,hko95}.

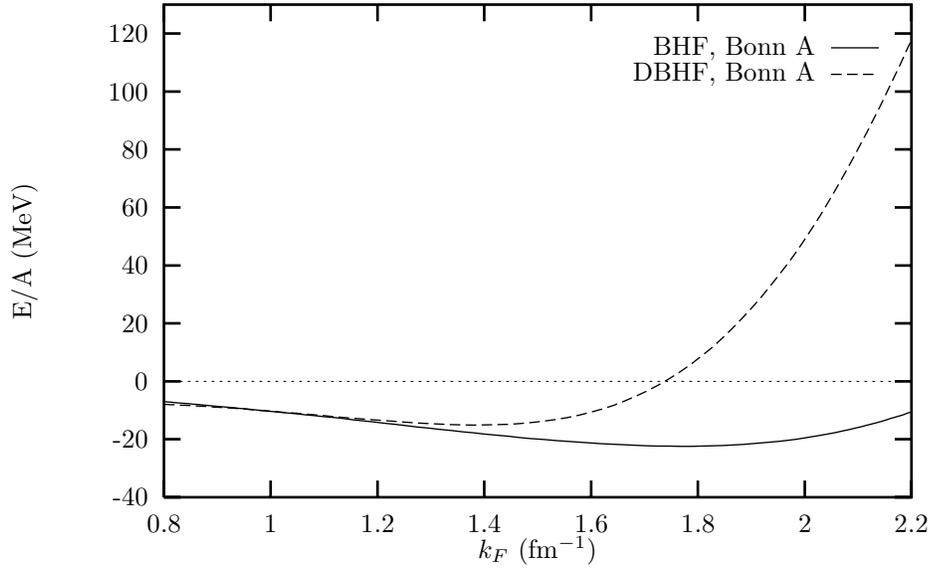
\begin{figure}
    % GNUPLOT: LaTeX picture with Postscript
\setlength{\unitlength}{0.1bp}
\special{!
%!PS-Adobe-2.0
%%Creator: gnuplot
%%DocumentFonts: Helvetica
%%BoundingBox: 50 50 770 554
%%Pages: (atend)
%%EndComments
/gnudict 40 dict def
gnudict begin
/Color false def
/Solid false def
/gnulinewidth 5.000 def
/vshift -33 def
/dl {10 mul} def
/hpt 31.5 def
/vpt 31.5 def
/M {moveto} bind def
/L {lineto} bind def
/R {rmoveto} bind def
/V {rlineto} bind def
/vpt2 vpt 2 mul def
/hpt2 hpt 2 mul def
/Lshow { currentpoint stroke M
  0 vshift R show } def
/Rshow { currentpoint stroke M
  dup stringwidth pop neg vshift R show } def
/Cshow { currentpoint stroke M
  dup stringwidth pop -2 div vshift R show } def
/DL { Color {setrgbcolor Solid {pop []} if 0 setdash }
 {pop pop pop Solid {pop []} if 0 setdash} ifelse } def
/BL { stroke gnulinewidth 2 mul setlinewidth } def
/AL { stroke gnulinewidth 2 div setlinewidth } def
/PL { stroke gnulinewidth setlinewidth } def
/LTb { BL [] 0 0 0 DL } def
/LTa { AL [1 dl 2 dl] 0 setdash 0 0 0 setrgbcolor } def
/LT0 { PL [] 0 1 0 DL } def
/LT1 { PL [4 dl 2 dl] 0 0 1 DL } def
/LT2 { PL [2 dl 3 dl] 1 0 0 DL } def
/LT3 { PL [1 dl 1.5 dl] 1 0 1 DL } def
/LT4 { PL [5 dl 2 dl 1 dl 2 dl] 0 1 1 DL } def
/LT5 { PL [4 dl 3 dl 1 dl 3 dl] 1 1 0 DL } def
/LT6 { PL [2 dl 2 dl 2 dl 4 dl] 0 0 0 DL } def
/LT7 { PL [2 dl 2 dl 2 dl 2 dl 2 dl 4 dl] 1 0.3 0 DL } def
/LT8 { PL [2 dl 2 dl 2 dl 2 dl 2 dl 2 dl 2 dl 4 dl] 0.5 0.5 0.5 DL } def
/P { stroke [] 0 setdash
  currentlinewidth 2 div sub M
  0 currentlinewidth V stroke } def
/D { stroke [] 0 setdash 2 copy vpt add M
  hpt neg vpt neg V hpt vpt neg V
  hpt vpt V hpt neg vpt V closepath stroke
  P } def
/A { stroke [] 0 setdash vpt sub M 0 vpt2 V
  currentpoint stroke M
  hpt neg vpt neg R hpt2 0 V stroke
  } def
/B { stroke [] 0 setdash 2 copy exch hpt sub exch vpt add M
  0 vpt2 neg V hpt2 0 V 0 vpt2 V
  hpt2 neg 0 V closepath stroke
  P } def
/C { stroke [] 0 setdash exch hpt sub exch vpt add M
  hpt2 vpt2 neg V currentpoint stroke M
  hpt2 neg 0 R hpt2 vpt2 V stroke } def
/T { stroke [] 0 setdash 2 copy vpt 1.12 mul add M
  hpt neg vpt -1.62 mul V
  hpt 2 mul 0 V
  hpt neg vpt 1.62 mul V closepath stroke
  P  } def
/S { 2 copy A C} def
end
}
\begin{picture}(3600,2160)(0,0)
\special{"
gnudict begin
gsave
50 50 translate
0.100 0.100 scale
0 setgray
/Helvetica findfont 100 scalefont setfont
newpath
-500.000000 -500.000000 translate
LTa
600 688 M
2817 0 V
LTb
600 251 M
63 0 V
2754 0 R
-63 0 V
600 470 M
63 0 V
2754 0 R
-63 0 V
600 688 M
63 0 V
2754 0 R
-63 0 V
600 907 M
63 0 V
2754 0 R
-63 0 V
600 1125 M
63 0 V
2754 0 R
-63 0 V
600 1344 M
63 0 V
2754 0 R
-63 0 V
600 1563 M
63 0 V
2754 0 R
-63 0 V
600 1781 M
63 0 V
2754 0 R
-63 0 V
600 2000 M
63 0 V
2754 0 R
-63 0 V
600 251 M
0 63 V
0 1795 R
0 -63 V
1002 251 M
0 63 V
0 1795 R
0 -63 V
1405 251 M
0 63 V
0 1795 R
0 -63 V
1807 251 M
0 63 V
0 1795 R
0 -63 V
2210 251 M
0 63 V
0 1795 R
0 -63 V
2612 251 M
0 63 V
0 1795 R
0 -63 V
3015 251 M
0 63 V
0 1795 R
0 -63 V
3417 251 M
0 63 V
0 1795 R
0 -63 V
600 251 M
2817 0 V
0 1858 V
-2817 0 V
600 251 L
LT0
3114 1946 M
180 0 V
600 611 M
2 0 V
3 0 V
5 0 V
6 -1 V
7 0 V
8 -1 V
10 -1 V
11 -1 V
13 -1 V
14 -1 V
15 -1 V
16 -2 V
18 -1 V
19 -2 V
20 -2 V
22 -2 V
22 -2 V
24 -2 V
25 -2 V
26 -3 V
27 -2 V
28 -3 V
30 -3 V
30 -3 V
31 -3 V
32 -3 V
34 -3 V
34 -4 V
34 -3 V
36 -4 V
37 -3 V
37 -4 V
38 -5 V
38 -4 V
40 -4 V
40 -5 V
40 -4 V
41 -5 V
41 -5 V
42 -4 V
43 -5 V
42 -4 V
43 -5 V
44 -4 V
43 -5 V
44 -4 V
44 -4 V
43 -5 V
44 -4 V
45 -3 V
44 -4 V
43 -3 V
44 -3 V
44 -3 V
43 -3 V
44 -2 V
43 -2 V
42 -2 V
43 -2 V
42 -1 V
41 0 V
41 -1 V
40 0 V
40 1 V
40 1 V
38 1 V
38 2 V
37 2 V
37 3 V
36 3 V
34 3 V
34 3 V
34 4 V
32 4 V
31 5 V
30 5 V
30 5 V
28 5 V
27 5 V
26 6 V
25 5 V
24 6 V
22 5 V
22 6 V
20 5 V
19 5 V
18 5 V
16 5 V
15 5 V
14 4 V
13 4 V
11 3 V
10 3 V
8 3 V
7 3 V
6 2 V
5 1 V
3 1 V
2 1 V
LT1
3114 1846 M
180 0 V
600 601 M
2 0 V
3 0 V
5 0 V
6 0 V
7 -1 V
8 0 V
10 0 V
11 -1 V
13 -1 V
14 0 V
15 -1 V
16 -1 V
18 -1 V
19 -1 V
20 -1 V
22 -2 V
22 -1 V
24 -2 V
25 -1 V
26 -2 V
27 -2 V
28 -3 V
30 -2 V
30 -2 V
31 -3 V
32 -3 V
34 -2 V
34 -3 V
34 -3 V
36 -3 V
37 -4 V
37 -3 V
38 -3 V
38 -3 V
40 -3 V
40 -3 V
40 -3 V
41 -3 V
41 -3 V
42 -2 V
43 -2 V
42 -2 V
43 -1 V
44 -1 V
43 1 V
44 1 V
44 1 V
43 3 V
44 4 V
45 5 V
44 6 V
43 7 V
44 9 V
44 11 V
43 12 V
44 14 V
43 17 V
42 18 V
43 20 V
42 22 V
41 24 V
41 26 V
40 28 V
40 31 V
40 32 V
38 35 V
38 37 V
37 38 V
37 40 V
36 41 V
34 42 V
34 43 V
34 44 V
32 44 V
31 46 V
30 46 V
30 47 V
28 47 V
27 46 V
26 46 V
25 45 V
24 44 V
22 42 V
22 41 V
20 40 V
19 37 V
18 36 V
16 34 V
15 31 V
14 29 V
13 27 V
11 23 V
10 22 V
8 18 V
7 16 V
6 12 V
5 10 V
3 7 V
2 3 V
stroke
grestore
end
showpage
}
\put(3054,1846){\makebox(0,0)[r]{DBHF, Bonn A}}
\put(3054,1946){\makebox(0,0)[r]{BHF, Bonn A}}
\put(2008,51){\makebox(0,0){$k_F$ (fm$^{-1}$)}}
\put(100,1180){%
\special{ps: gsave currentpoint currentpoint translate
270 rotate neg exch neg exch translate}%
\makebox(0,0)[b]{\shortstack{E/A (MeV)}}%
\special{ps: currentpoint grestore moveto}%
}
\put(3417,151){\makebox(0,0){2.2}}
\put(3015,151){\makebox(0,0){2}}
\put(2612,151){\makebox(0,0){1.8}}
\put(2210,151){\makebox(0,0){1.6}}
\put(1807,151){\makebox(0,0){1.4}}
\put(1405,151){\makebox(0,0){1.2}}
\put(1002,151){\makebox(0,0){1}}
\put(600,151){\makebox(0,0){0.8}}
\put(540,2000){\makebox(0,0)[r]{120}}
\put(540,1781){\makebox(0,0)[r]{100}}
\put(540,1563){\makebox(0,0)[r]{80}}
\put(540,1344){\makebox(0,0)[r]{60}}
\put(540,1125){\makebox(0,0)[r]{40}}
\put(540,907){\makebox(0,0)[r]{20}}
\put(540,688){\makebox(0,0)[r]{0}}
\put(540,470){\makebox(0,0)[r]{-20}}
\put(540,251){\makebox(0,0)[r]{-40}}
\end{picture}
    \caption{EOS for symmetric nuclear matter with the NN potentials 
 and many-body methods described in the text.}
    \label{fig:fig1}
\end{figure}

The DBHF is a variational procedure where the single-particle
energies are obtained through an iterative self-consistency scheme.
To obtain the relativistic single-particle energies
we solve the Dirac equation for
a nucleon in the nuclear 
medium, with $c=\hbar=1$,
\begin{equation}
       (\not p -m +\Sigma (p))\tilde{u}(p,s)=0,
\end{equation}
where $m$ is the free nucleon mass and $\tilde{u}(p,s)$ is  
the Dirac spinor for positive energy solutions, 
 $p=(p^0 ,{\bf p})$ being
a four momentum and  $s$ the spin projection.
The self-energy $\Sigma (p)$ 
for nucleons can be written as
\begin{equation}
       \Sigma(p) =
       \Sigma_S(p) -\gamma_0 \Sigma^0(p)
       +\mbox{\boldmath $\gamma$}\cdot {\bf p}\Sigma^V(p).
\end{equation}
Since $\Sigma^V << 1$ \cite{brock90,sw86}, we approximate 
the self-energy by
\begin{equation}
       \Sigma \approx \Sigma_S -\gamma_0 \Sigma^0 = U_S + U_V,
\end{equation}
where $U_S$ is an attractive
scalar field and $U_V$ is the time-like component
of a repulsive vector field.
The Dirac spinor then reads  
\begin{equation}
       \tilde{u}(p,s)=\sqrt{\frac{\tilde{E}_p+\tilde{m}}{2\tilde{m}}}
       \left(\begin{array}{c} \chi_s\\ \\
       \frac{\mbox{\boldmath $\sigma$}\cdot{\bf p}}
       {\tilde{E}_p+\tilde{m}}\chi_s
       \end{array}\right),
\end{equation}
where 
$\chi_s$ is the Pauli
spinor and 
terms with tilde like $\tilde{E}_p=\sqrt{{\bf p}^2+\tilde{m}^2}$
represent medium modified quantities.
Here we have defined \cite{brock90,sw86} $\tilde{m}=m+U_S$.
The single-particle energies 
$\tilde{\varepsilon}_p$ can then be written as
\begin{equation}
       \tilde{\varepsilon}_p=
       \left\langle  p\right|\mbox{\boldmath $\gamma$}\cdot {\bf p}
        +m\left| p \right\rangle+u_p=
       \tilde{E}_{\bf p} +U_V,
       \label{eq:sprelen}
\end{equation}
where the single-particle potential $u_p$ is given by 
$u_p=U_S\tilde{m}/\tilde{E}_p+U_V$ and 
can in turn be defined in terms
of the $G$-matrix
\begin{equation}
       u_p =\sum_{p'\leq k_F} \frac{\tilde{m}^2}{\tilde{E}_{{\bf p}'}
       \tilde{E}_p}
       \left\langle  pp' \right|
       G(\omega =\tilde{\varepsilon}_p
       +\tilde{\varepsilon}_p') \left| pp' \right\rangle,
       \label{eq:urel}
\end{equation}
where $p,p'$ represent quantum numbers like momentum, spin, isospin projection
etc of the different single-particle states and $k_F$ is the Fermi
momentum. 
Eqs.\ (\ref{eq:sprelen})-(\ref{eq:urel}) are solved self-consistently
starting
with adequate values for the scalar and vector components
$U_S$ and $U_V$.  The energy per particle can then be calculated from 
\begin{eqnarray}
   \frac{{\cal E}}{A}&=&\frac{1}{A}\sum_{p'\leq k_F}\frac{\tilde{m}m+{\bf p'}
^2}{\tilde{E}_{{\bf p}'}} \nonumber \\
&+&\frac{1}{2A}\sum_{p'p''\leq k_F}
\frac{\tilde{m}^2\langle p'p''|G(\tilde{E}=\tilde{\epsilon}_p'+
\tilde{\epsilon}_{p''})|p'p''\rangle _{\rm AS}}
{\tilde{E}_{{\bf p}'}\tilde{E}_{{\bf p}''}}
-m.
\label{eq:energrel}
\end{eqnarray}
In Fig. \ref{fig:fig1} we show the EOS obtained in our non-relativistic 
and relativistic calculations.   The non-relativistic one fails to 
meet the empirical data, while the relativistic calculation very nearly 
succeeds.

Having obtained in-medium single-particle energies, we proceed to solve 
the coupled gap equations for $^3S_1$-$^3D_1$ pairing.  
Employing an angle-average approximation, these can be written 
\cite{tak93}
\begin{eqnarray}
\Delta_0(k)&=&-\int_0^{\infty}dk'k'^2 \frac{1}{E(k')} 
 (V_{00}(k,k')\Delta_0(k')   \nonumber \\
 & & -V_{02}(k,k')\Delta_2(k')) 
\label{eq:gap1}
\end{eqnarray}
\begin{eqnarray}
\Delta_2(k)&=&-\int_0^{\infty}dk'k'^2 \frac{1}{E(k')} 
 (-V_{20}(k,k')\Delta_0(k') \nonumber \\
 & & +V_{22}(k,k')\Delta_2(k')) 
\label{eq:gap2}
\end{eqnarray}   
where the subscripts $0$ and $2$ denote $S$ and $D$ states respectively, 
$V_{ll'}$ is the free momentum-space NN interaction in the relevant channel, 
$\Delta_0$ and $\Delta_2$ are the $S$ and $D$ state gap functions 
respectively, and $E(k)$ is the quasi-particle energy given by 
\begin{equation}
E(k)=\sqrt{(\epsilon_k-\mu)^2+\Delta_0(k)^2+\Delta_2(k)^2},  
\label{eq:gap3}
\end{equation}
where $\mu$ is the chemical potential. 
The quantity 
\begin{equation}
D_F=\sqrt{\Delta_0(k_F)^2+\Delta_2(k_F)^2}
\label{eq:gap4}
\end{equation}
will in the following be referred to as the energy gap, according to 
the conventional definition \cite{alm90,vonder91,tak93,baldo95}.   
For $^3S_1$-$^3D_1$ pairing it is also necessary to solve the 
equation for particle number conservation 
\begin{equation}
\rho \equiv \frac{2k_F^3}{3\pi^2}=\frac{1}{\pi^2}\int_{0}^{\infty}dk k^2 
\left( 1-\frac{\epsilon_k-\mu}{E(k)}\right )
\label{eq:gap5}
\end{equation}
for $\mu$ self-consistently together with Eqs.\ (\ref{eq:gap1}) and 
(\ref{eq:gap2}).  

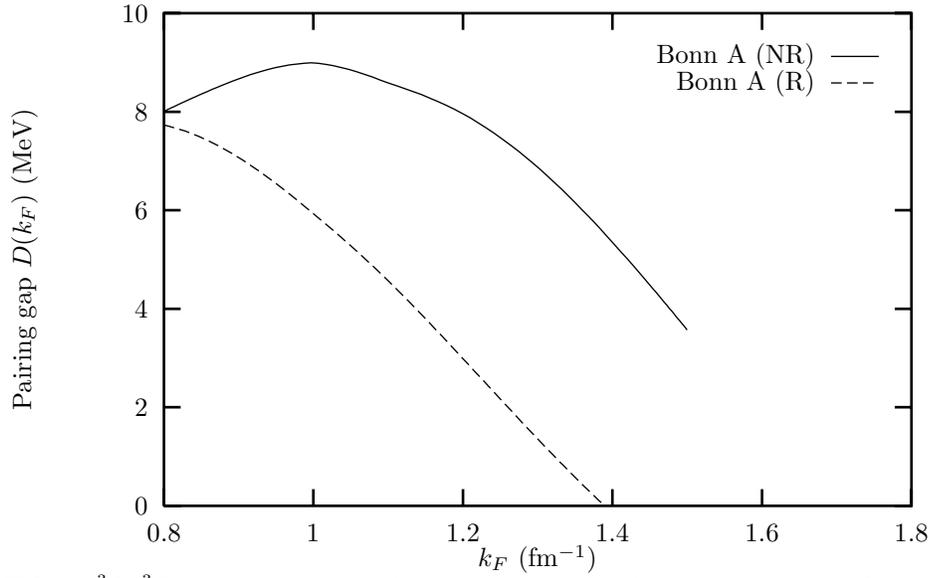
\begin{figure}
    % GNUPLOT: LaTeX picture with Postscript
\setlength{\unitlength}{0.1bp}
\special{!
%!PS-Adobe-2.0
%%Creator: gnuplot
%%DocumentFonts: Helvetica
%%BoundingBox: 50 50 770 554
%%Pages: (atend)
%%EndComments
/gnudict 40 dict def
gnudict begin
/Color false def
/Solid false def
/gnulinewidth 5.000 def
/vshift -33 def
/dl {10 mul} def
/hpt 31.5 def
/vpt 31.5 def
/M {moveto} bind def
/L {lineto} bind def
/R {rmoveto} bind def
/V {rlineto} bind def
/vpt2 vpt 2 mul def
/hpt2 hpt 2 mul def
/Lshow { currentpoint stroke M
  0 vshift R show } def
/Rshow { currentpoint stroke M
  dup stringwidth pop neg vshift R show } def
/Cshow { currentpoint stroke M
  dup stringwidth pop -2 div vshift R show } def
/DL { Color {setrgbcolor Solid {pop []} if 0 setdash }
 {pop pop pop Solid {pop []} if 0 setdash} ifelse } def
/BL { stroke gnulinewidth 2 mul setlinewidth } def
/AL { stroke gnulinewidth 2 div setlinewidth } def
/PL { stroke gnulinewidth setlinewidth } def
/LTb { BL [] 0 0 0 DL } def
/LTa { AL [1 dl 2 dl] 0 setdash 0 0 0 setrgbcolor } def
/LT0 { PL [] 0 1 0 DL } def
/LT1 { PL [4 dl 2 dl] 0 0 1 DL } def
/LT2 { PL [2 dl 3 dl] 1 0 0 DL } def
/LT3 { PL [1 dl 1.5 dl] 1 0 1 DL } def
/LT4 { PL [5 dl 2 dl 1 dl 2 dl] 0 1 1 DL } def
/LT5 { PL [4 dl 3 dl 1 dl 3 dl] 1 1 0 DL } def
/LT6 { PL [2 dl 2 dl 2 dl 4 dl] 0 0 0 DL } def
/LT7 { PL [2 dl 2 dl 2 dl 2 dl 2 dl 4 dl] 1 0.3 0 DL } def
/LT8 { PL [2 dl 2 dl 2 dl 2 dl 2 dl 2 dl 2 dl 4 dl] 0.5 0.5 0.5 DL } def
/P { stroke [] 0 setdash
  currentlinewidth 2 div sub M
  0 currentlinewidth V stroke } def
/D { stroke [] 0 setdash 2 copy vpt add M
  hpt neg vpt neg V hpt vpt neg V
  hpt vpt V hpt neg vpt V closepath stroke
  P } def
/A { stroke [] 0 setdash vpt sub M 0 vpt2 V
  currentpoint stroke M
  hpt neg vpt neg R hpt2 0 V stroke
  } def
/B { stroke [] 0 setdash 2 copy exch hpt sub exch vpt add M
  0 vpt2 neg V hpt2 0 V 0 vpt2 V
  hpt2 neg 0 V closepath stroke
  P } def
/C { stroke [] 0 setdash exch hpt sub exch vpt add M
  hpt2 vpt2 neg V currentpoint stroke M
  hpt2 neg 0 R hpt2 vpt2 V stroke } def
/T { stroke [] 0 setdash 2 copy vpt 1.12 mul add M
  hpt neg vpt -1.62 mul V
  hpt 2 mul 0 V
  hpt neg vpt 1.62 mul V closepath stroke
  P  } def
/S { 2 copy A C} def
end
}
\begin{picture}(3600,2160)(0,0)
\special{"
gnudict begin
gsave
50 50 translate
0.100 0.100 scale
0 setgray
/Helvetica findfont 100 scalefont setfont
newpath
-500.000000 -500.000000 translate
LTa
600 251 M
2817 0 V
LTb
600 251 M
63 0 V
2754 0 R
-63 0 V
600 623 M
63 0 V
2754 0 R
-63 0 V
600 994 M
63 0 V
2754 0 R
-63 0 V
600 1366 M
63 0 V
2754 0 R
-63 0 V
600 1737 M
63 0 V
2754 0 R
-63 0 V
600 2109 M
63 0 V
2754 0 R
-63 0 V
600 251 M
0 63 V
0 1795 R
0 -63 V
1163 251 M
0 63 V
0 1795 R
0 -63 V
1727 251 M
0 63 V
0 1795 R
0 -63 V
2290 251 M
0 63 V
0 1795 R
0 -63 V
2854 251 M
0 63 V
0 1795 R
0 -63 V
3417 251 M
0 63 V
0 1795 R
0 -63 V
600 251 M
2817 0 V
0 1858 V
-2817 0 V
600 251 L
LT0
3114 1946 M
180 0 V
600 1739 M
1 1 V
3 1 V
3 1 V
4 2 V
5 3 V
6 2 V
7 4 V
8 3 V
8 4 V
10 5 V
11 5 V
11 5 V
13 6 V
13 6 V
14 6 V
15 7 V
16 7 V
17 7 V
17 8 V
18 7 V
19 8 V
20 8 V
20 8 V
22 8 V
22 8 V
22 8 V
23 7 V
24 7 V
25 7 V
24 6 V
26 5 V
26 5 V
27 3 V
27 3 V
27 2 V
28 -2 V
28 -4 V
29 -5 V
29 -7 V
29 -7 V
30 -9 V
30 -10 V
30 -10 V
30 -12 V
31 -11 V
30 -10 V
31 -11 V
31 -11 V
31 -11 V
30 -13 V
31 -13 V
31 -14 V
31 -15 V
30 -16 V
31 -17 V
30 -19 V
30 -19 V
30 -20 V
29 -21 V
30 -22 V
29 -22 V
28 -23 V
29 -24 V
28 -24 V
27 -25 V
27 -25 V
27 -25 V
26 -26 V
25 -25 V
25 -26 V
25 -25 V
24 -25 V
23 -25 V
22 -25 V
22 -24 V
21 -24 V
21 -23 V
20 -22 V
19 -22 V
18 -21 V
17 -20 V
17 -20 V
16 -19 V
15 -17 V
14 -17 V
13 -16 V
13 -15 V
11 -14 V
11 -12 V
9 -12 V
9 -11 V
8 -9 V
7 -9 V
6 -7 V
5 -6 V
4 -5 V
3 -4 V
2 -3 V
2 -1 V
LT1
3114 1846 M
180 0 V
600 1687 M
1 0 V
2 -1 V
3 0 V
3 -1 V
4 -1 V
5 -1 V
6 -2 V
7 -2 V
7 -2 V
8 -2 V
9 -3 V
10 -3 V
11 -3 V
11 -4 V
12 -5 V
12 -5 V
14 -5 V
14 -6 V
14 -7 V
16 -7 V
16 -8 V
16 -8 V
18 -10 V
18 -10 V
18 -10 V
19 -12 V
20 -12 V
20 -13 V
20 -14 V
21 -14 V
22 -16 V
22 -16 V
22 -16 V
23 -18 V
23 -18 V
24 -19 V
24 -20 V
24 -20 V
24 -20 V
25 -21 V
25 -21 V
25 -21 V
25 -22 V
26 -23 V
26 -23 V
25 -23 V
26 -24 V
26 -24 V
26 -25 V
26 -25 V
26 -26 V
26 -26 V
26 -26 V
26 -27 V
25 -26 V
26 -27 V
25 -27 V
25 -26 V
25 -27 V
25 -27 V
24 -26 V
25 -26 V
23 -26 V
24 -25 V
23 -25 V
23 -25 V
22 -24 V
22 -24 V
22 -23 V
21 -23 V
20 -22 V
20 -21 V
20 -21 V
19 -20 V
18 -19 V
18 -19 V
18 -18 V
16 -17 V
16 -17 V
16 -15 V
14 -15 V
14 -14 V
14 -14 V
12 -12 V
12 -12 V
11 -11 V
11 -10 V
10 -10 V
9 -8 V
8 -8 V
7 -7 V
7 -7 V
6 -5 V
5 -5 V
4 -4 V
3 -3 V
3 -3 V
stroke
grestore
end
showpage
}
\put(3054,1846){\makebox(0,0)[r]{Bonn A (R)}}
\put(3054,1946){\makebox(0,0)[r]{Bonn A (NR)}}
\put(2008,51){\makebox(0,0){$k_F$ (fm$^{-1}$)}}
\put(100,1180){%
\special{ps: gsave currentpoint currentpoint translate
270 rotate neg exch neg exch translate}%
\makebox(0,0)[b]{\shortstack{Pairing gap $D(k_F)$ (MeV)}}%
\special{ps: currentpoint grestore moveto}%
}
\put(3417,151){\makebox(0,0){1.8}}
\put(2854,151){\makebox(0,0){1.6}}
\put(2290,151){\makebox(0,0){1.4}}
\put(1727,151){\makebox(0,0){1.2}}
\put(1163,151){\makebox(0,0){1}}
\put(600,151){\makebox(0,0){0.8}}
\put(540,2109){\makebox(0,0)[r]{10}}
\put(540,1737){\makebox(0,0)[r]{8}}
\put(540,1366){\makebox(0,0)[r]{6}}
\put(540,994){\makebox(0,0)[r]{4}}
\put(540,623){\makebox(0,0)[r]{2}}
\put(540,251){\makebox(0,0)[r]{0}}
\end{picture}
    \caption{$^3S_1$-$^3D_1$ energy gap in nuclear matter, calculated in 
  non-relativistic (full and long-dashed lines) and relativistic 
 (short-dashed line) approaches.  }
    \label{fig:fig2}
\end{figure}

In the non-relativistic calculation, we have used  
the Bonn A potential with parameters from table A.1 in Ref.\ \cite{mac89}.  
The results are shown in 
Fig.\ \ref{fig:fig2}.   We found a 
large energy gap at the empirical saturation density, around $6$ MeV 
at $k_F=1.36\;{\rm fm}^{-1}$, in agreement with  
earlier non-relativistic calculations 
\cite{alm90,vonder91,tak93,baldo95}.  

Recently, several groups have developed relativistic formulations of 
pairing in nuclear matter \cite{ring91,guim96,matera97}, and have   
applied them to $^1S_0$ pairing.  
The models are of the Walecka-type \cite{sw86} in the sense that 
meson masses and coupling constants are fitted 
so that the mean-field EOS of nuclear 
matter meets the empirical data.  In this way, however, the relation of 
the models to free-space NN scattering becomes somewhat unclear.  
An interesting result found in Refs.\ \cite{ring91,guim96,matera97} 
is that the $^1S_0$ energy gap vanishes at densities slightly below the 
empirical saturation density.  This is in contrast with 
non-relativistic calculations which generally give a relatively small, 
but non-vanishing $^1S_0$ gap at this density, see for instance  
\cite{kuch89,baldo90,chen93,elg961}.   

With these results in mind, we found it interesting to consider 
relativistic effects on $^3S_1$-$^3D_1$ pairing in nuclear matter, 
which, to our knowledge, has not been done before.  
A simple way of doing this is to incorporate 
minimal relativity in the gap equation, thus using DBHF single-particle 
energies in the energy denominators and modifying the free NN interaction 
by a factor $\tilde{m}^2/\tilde{E}_k\tilde{E}_{k'}$ \cite{elg962}.  
With this prescription, we obtained the results shown in Fig.\ 
\ref{fig:fig2} (short-dashed line).  As can be seen, the gap at the 
empirical saturation density is reduced from 6 MeV to nearly zero.  

Let us try to understand the difference between the 
relativistic and the non-relativistic calculation.  First of all, 
we point out that it is well known that the introduction of relativity 
in the many-body problem leads to increased repulsion at densities 
at and above the saturation density \cite{brown87}.  Even a slight 
increase in the repulsion might have large consequences for the energy 
gap, as the gap depends exponentially on the interaction at the 
Fermi surface.  One can obtain a numerical estimate of 
the effect as follows.  If one takes the weak-coupling limit 
of the gap equations, Eqs. (\ref{eq:gap1}) and (\ref{eq:gap2}), 
it is easy to show that one obtains the same form for the 
energy gap as for $^1S_0$ pairing,  
\begin{equation}
 D_F=2\delta\epsilon\exp\left[-\frac{1}{N(k_F)V_{pair}}\right], 
\label{eq:gap6}
\end{equation}
where $\delta\epsilon$ is an appropriate energy interval, 
$N(k_F)=\Omega m^* k_F/2\pi^2\hbar^2$ is the 
density of states at the Fermi surface, $m^*$ is the 
nucleon effective mass, and $\Omega$ is the 
volume occupied by the system.   
The pairing interaction at the Fermi surface, $V_{pair}$, is given by  
\begin{equation}
V_{pair}=\frac{\sqrt{(V_{SS}-V_{DD})^2+4V_{SD}^2}-V_{SS}
-V_{SD}}{2}, 
\label{eq:gap7}
\end{equation}
where $V_{SS}=V_{00}(k_F,k_F)$, $V_{SD}=V_{02}(k_F,k_F)=V_{20}(k_F,k_F)$, 
and $V_{DD}=V_{22}(k_F,k_F)$.    
With our relativistic approach to the gap equation, the corresponding 
weak-coupling expression for the gap is obtained by replacing 
$V_{pair}$ with $\tilde{m}^2/\tilde{E}^2_{k_F}V_{pair}$ and using 
the relativistic expression for the density of states instead of the 
non-relativistic one.  We will now consider the saturation density, 
and take $k_F\approx 1.4\;{\rm fm}^{-1}$.  
The non-relativistic single-particle spectrum was parameterized as 
$\epsilon_k=k^2/2m^*+U_0$.  At $k_F=1.4\;{\rm fm}^{-1}$ the values 
of the parameters were $m^{*}/m=0.6751$, $U_0=-97.2755$ MeV.  
For the relativistic single-particle spectrum the relevant quantities 
were $U_S=-384.89$ MeV, $U_V=300.18$ MeV.  
We found $N_R(k_F=1.4\;{rm fm}^{-1})/N_{NR}(k_F=1.4\;
{\rm fm}^{-1})\approx 1$, and then 
\begin{eqnarray}
\frac{D_F^{R}}{D_F^{NR}}&=&\exp\left[-\frac{1}{N_{NR}(1.4)V_{pair}}
\left(\frac{\tilde{E}^2_F}{\tilde{m}^2}-1\right)\right] \nonumber \\
 &\approx&\left(\frac{D_F^{NR}}{2\delta\epsilon}\right)^{1/4},  \nonumber
\end{eqnarray}  
where the superscript $R$ refers to relativistic quantities, $NR$ to 
non-relativistic ones.
If we make the common choice $\delta\epsilon=\epsilon^{NR}_F$ and use 
$D_F\approx 6$ MeV, we obtain 
\[
\frac{D_F^R}{D_F^{NR}}\approx 0.5, 
\]
thus, the introduction of relativity in the gap equation suppresses the 
gap at the saturation density by a factor of two.  This argument 
makes it reasonable that relativistic effects reduce the gap.  That 
the reduction is larger in the full calculation than in this simple 
estimate is understandable, since we in the weak coupling approximation 
neglect the momentum-dependence of the interaction.  More specifically, 
the repulsive high-momentum components are left out, and these will 
reduce the gap further.  

It is also interesting to obtain the ratio $D_F^{R}/D_F^{NR}$ at 
the respective saturation densities for the relativistic and 
non-relativistic calculations.  
The non-relativistic EOS saturates at $k_F\approx 1.8\;{\rm fm}^{-1}$. 
At this density, we had numerical problems with solving the gap 
equations, something which may occur when the gap is small.  
However, using the weak-coupling expression for the gap, we could  
estimate 
$D_F^{R}(k_F=1.4\;{\rm fm}^{-1})/D_F^{NR}(k_F=1.8\;{\rm fm}^{-1})$.  
First we used the non-relativistic gaps in the density range 
$k_F=1.2$-$1.4\;{\rm fm}^{-1}$ to calculate 
$N_{NR}(k_F)V_{pair}(k_F)$, and fitted the results with a quadratic 
polynomial in $k_F$.  From this fit we estimated $N_{NR}(k_F=1.8)
V_{pair}(k_F=1.8)\approx 0.257$.  Then 
\begin{eqnarray}
 \frac{N_{NR}(1.8)V_{pair}(1.8)}{N_{NR}(1.4)V_{pair}(1.4)}&=&
\frac{0.257}{0.333} \nonumber \\ 
\Rightarrow V_{pair}(1.8)&=&\frac{N_{NR}(1.4)}{N_{NR}(1.8)}
\frac{0.257}{0.333}V_{pair}(1.4) \nonumber \\
 &\approx&0.695V_{pair}(1.4) \nonumber
\end{eqnarray}
where we have used $m^*(1.4)/m=0.675$, $m^*(1.8)/m=0.5834$ in the 
densities of states.  We then formed the ratio $D_F^R(1.4)/D_F^{NR}(1.8)$ 
and obtained 
\begin{eqnarray}
\frac{D_F^R}{D_F^{NR}}&=&\exp[-\frac{1}{N_{NR}(1.8)V_{pair}(1.4)} 
 \nonumber \\  
 &\times&  \left(\frac{N_{NR}(1.8)}{N_R(1.4)}
 \frac{\tilde{E}_F^2(1.4)}{\tilde{m}^2(1.4)} 
-\frac{1}{0.695}\right)] \nonumber \\ 
 &\approx&1, \nonumber 
\end{eqnarray}
since we found $N_{NR}(1.8)/N_{R}(1.4)\approx 1.15$, 
$\tilde{E}_F^2(1.4)/\tilde{m}^2(1.4)\approx 1.25$, thus making the 
expression in the inner parenthesis $\approx 0$.   
Although this argument is only indicative, it makes it reasonable 
to assume that the non-relativistic gap will be very small at the 
calculated non-relativistic saturation density.   

In this Letter, we have presented non-relativistic and relativistic 
calculations of the $^3S_1$-$^3D_1$ np gap in symmetric nuclear matter.  
The non-relativistic calculations gives a large gap of approximately 
6 MeV at the empirical saturation density.  In the relativistic calculation 
we find that the gap is vanishingly small  at this density.  
This is the main result 
of this Letter.  Non-relativistic calculations with two-body 
interactions will in general give a saturation density which is too high, 
an example of which is shown in Fig. \ref{fig:fig1}.  Thus in the 
non-relativistic approach we are actually calculating the 
gap at density below the theoretical saturation density, and one 
may question the physical relevance of a large gap at a density where 
the system is theoretically unstable.  If one looks at the gap 
at the {\em calculated} saturation density, it is in fact close 
to zero.  
In the DBHF calculation we come 
very close to reproducing the empirical saturation density and binding 
energy, and when this is used as a starting point for a BCS calculation, 
we find that the gap vanishes, both at the empirical and the calculated 
saturation density.  That the DBHF calculation meets the empirical 
points is perhaps fortuitous, as important many-body diagrams are 
neglected and only medium modifications of the nucleon mass are 
accounted for.  However, a recent investigation where medium 
modifications of meson masses were included, 
showed that the results of the DBHF do not change very much, and in 
particular the saturation properties are still very good \cite{rapp97}.
Nevertheless, the essential property which is needed in all 
non-relativistic models to 
get to the empirical point is an increased repulsion at and around the 
empirical saturation density.  Regardless of the mechanism, this may 
reduce the pairing gap dramatically.  The main point we wish to make 
in this Letter is thus that the inclusion of these additional repulsive 
effects may destroy pairing at the empirical saturation density, 
a result which is in accordance with the lack of empirical evidence 
for the strong np pairing produced in non-relativistic calculations.

\end{multicols}

\clearpage

\end{document}